\documentclass[12pt]{article}
\pdfoutput=1
\usepackage{amsmath,amssymb}
\usepackage{amsfonts}
\usepackage{setspace}
\usepackage{subcaption}
\usepackage{xcolor}
\usepackage{enumerate}
\usepackage{MnSymbol}
\usepackage{graphicx}
\usepackage[hidelinks]{hyperref} 
\numberwithin{equation}{section}
\usepackage[utf8]{inputenc}
\newcommand{\gsim}{\lower.7ex\hbox{$\;\stackrel{\textstyle>}{\sim}\;$}}
\newcommand{\lsim}{\lower.7ex\hbox{$\;\stackrel{\textstyle<}{\sim}\;$}}
\def\O{{\mathcal O}}

\newcommand{\be}{\begin{equation}}
\newcommand{\ee}{\end{equation}}
\newcommand{\bea}{\begin{eqnarray}}
\newcommand{\eea}{\end{eqnarray}}

\newcommand{\comment}[1]{}
\newcommand{\expect}[1]{\left\langle #1 \right\rangle}

\newcommand{\bsb}{\boldsymbol}
\newcommand{\Cint}{C\kern-1.1em\int}

\def\ep{\epsilon}

\def\d{\partial}
\def\vphi{\varphi}

\def\O{\mathcal{O}}

\def\r{{\bsb r}}

\def\S{{\mathcal{S}}}

\def\arctanh{{\rm arctanh}}

\def\Re{{\rm Re~}}

\def\hyp{{}_2 F_1}
\def\S{\mathcal{S}}
\begin{document}
\vspace*{-1. cm}
\begin{center}
{\bf \Large Probing de Sitter from the horizon}
\vskip 1cm
{{Mehrdad Mirbabayi$^a$} and {Flavio Riccardi$^{b,c}$}}
\vskip 0.5cm
       {\normalsize $^a${\em International Centre for Theoretical Physics, Trieste, Italy}}\\
       {\normalsize $^b${\em Scuola Internazionale di Studi Superiori Avanzati, Trieste, Italy}}\\
       {\normalsize $^c${\em I.N.F.N. sezione di Trieste, Trieste, Italy}}
\end{center}

\vspace{.8cm}
{\noindent \textbf{Abstract:}  
In a QFT on de Sitter background, one can study correlators between fields pushed to the future and past horizons of a comoving observer. This is a neat probe of the physics in the observer's causal diamond (known as the static patch). We use this observable to give a generalization of the quasinormal spectrum in interacting theories, and to connect it to the spectral density that appears in the K\"all\'en-Lehmann expansion of dS correlators. We also introduce a finite-temperature effective field theory consisting of free bulk fields coupled to a boundary. In matching it to the low frequency expansion of correlators, we find positivity constraints on the EFT parameters following from unitarity. 

\vspace{0.3cm}
\vspace{-1cm}
\vskip 1cm
\section{Introduction}
An important characteristic of a spacetime with a horizon is the spectrum of exponents that control the relaxation of perturbations. At linear level, these exponents coincide with the (generically complex) frequencies of the {\em quasinormal} modes \cite{Konoplya}. But they can also be thought of as the poles of the retarded Green's function.

De Sitter spacetime has observer-dependent horizons, and is described within these horizons by the metric 
\be
ds^2 = -(1-r^2) dt^2 + \frac{dr^2}{1-r^2}+ r^2 d\Omega^2.
\ee
It is therefore natural to explore the quasinormal-mode (QNM) spectrum of fields on dS \cite{Brady}. One motivation is that we believe there has been a long period of inflation in our past. This is a geometry that is locally close to dS, and the above-mentioned exponents control the power-law decay of cosmological correlators. dS is also a clean setup to learn about the relaxation process. For instance, the infrared dynamics of light scalar fields in de Sitter provides a rare example of a Markovian process whose decay exponents can be explicitly computed in terms of the UV couplings \cite{Markov}. 

An interesting feature of free field theories on dS background is the fact that the poles associated to the quasinormal modes are the only singularities of the retarded Green's function, while in asymptotically flat black hole spacetimes there are additional singularities (branch-points) corresponding to the power-law tails \cite{Ching}. One goal of this work is to explore the analytic structure of the retarded function at the interacting level. 

To this end, we find it convenient to consider in-out correlators obtained by pushing the operator insertions to the future and past horizons (see figure \ref{fig:S}). Operationally, this can be thought of as a scattering experiment in the static patch, where the scattering states are prepared by a family of accelerating observers at fixed $r$, in the limit $r\to 1$. We will show in section \ref{sec:free} that the poles of the retarded in-out correlators coincide with the well-known spectrum of quasinormal modes for free scalar fields. 
\begin{figure}[t]
\centering
  \includegraphics[scale=1.]{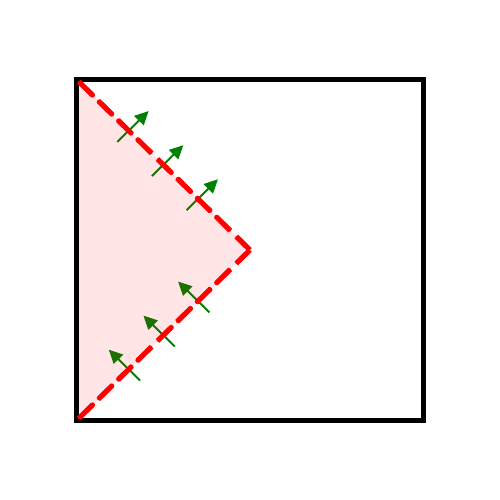}
  \caption{The Penrose diagram of dS. We study in and out states in one static patch (shaded).}
  \label{fig:S}
\end{figure}

At the interacting level, we use the K\"all\'en-Lehmann representation for the 2-point correlators in dS. That is, expand the interacting 2-point correlator in terms of correlators of free fields associated to the unitary representations of dS isometry group \cite{Bros}. Schematically 
\be\label{XY0}
\expect{\phi(X) \phi(Y)} = \sum_{\Delta} \rho(\Delta)
G_f(X,Y;\Delta).
\ee
We will see how the singularity structure of the retarded function (in $\omega$) is inherited from that of the spectral density $\rho(\Delta)$. Explicit examples and perturbative arguments given in \cite{Hogervorst,DiPietro} suggest that $\rho(\Delta)$ is {\em generically} a meromorphic function. This implies that the static patch retarded function too will generically have a set of isolated poles as its singularities. 

A priori, this sounds surprising.\footnote{We thank Victor Gorbenko for bringing this curious feature to our attention.} Correlators of interacting theories typically have a more intricate singularity structure (e.g. loops usually introduce branch-cuts). The finiteness of volume in the static patch does not seem to be a good explanation for the simplicity of its correlators. In fact, the static patch Hamiltonian ($i \d_t$) has a continuous spectrum, and one can find (rare) examples with a continuous spectrum of decay exponents. We present one in section \ref{sec:counter}.

In section \ref{sec:rindler}, we present an analogy that might be helpful in understanding why the spectrum is generically discrete. There we see that the Rindler-space correlator is a meromorphic function of Rindler frequency for a generic interacting theory on $2d$ Minkowski spacetime. 

Apart from the simple analytic structure, the spectral density $\rho(\Delta)$ was shown in \cite{Hogervorst,DiPietro} to be non-negative as a consequence of unitarity. It is often the case that the unitarity of the UV completion constrains the low energy observables and the effective field theory that governs them \cite{Adams}. In section \ref{sec:EFT}, we will find a similar application. Indeed, the small $\omega$ expansion of in-out correlators of a free theory can be reproduced by a {\em boundary effective field theory} (BEFT). As we will see, the emergence of this BEFT is quite natural when using the tortoise coordinate $x  = \arctanh \ r$, with $0\leq x< \infty$. Spherical modes of the field are effectively free $2d$ fields except in the vicinity of $0<x\sim 1$ (see figure \ref{fig:Veff}). For small $\omega$ this region can be collapsed to a one-dimensional boundary on which composite operators are localized. 
\begin{figure}[t]
\centering
  \includegraphics[scale=.8]{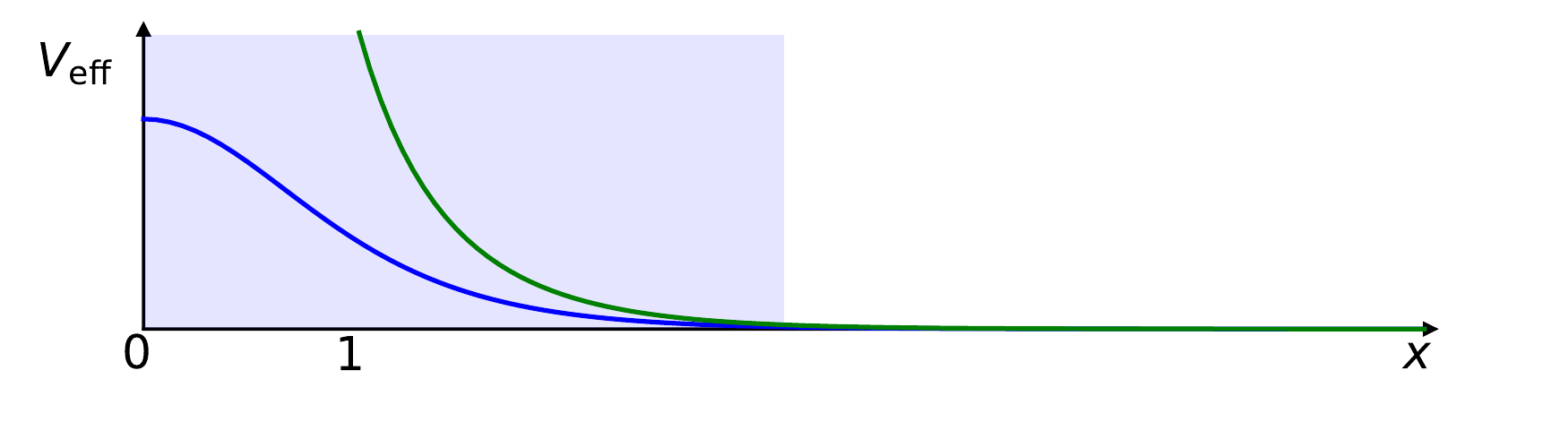}
  \caption{The effective potential for the spherical harmonics of a free field consists of a mass term (blue) and a centrifugal barrier (green). They decay exponentially toward the horizon $x\to \infty$. When $\omega \ll 1$, the interaction region (shaded) can be collapsed to a boundary with localized interactions.}
  \label{fig:Veff}
\end{figure}

In section \ref{sec:intBEFT}, we will discuss the condition under which a similar BEFT description is applicable to interacting theories. Naturally, this BEFT has to be formulated at finite temperature, and in matching it with \eqref{XY0} it is necessary to include dissipative terms. We will see that unitarity leads to positivity constraints on the EFT coefficients. We will conclude in section \ref{sec:con}.
\section{Quasinormal spectrum}
\subsection{In-out correlators}
When it is possible to define the scattering matrix, it is often a more convenient observable to work with than the local correlation functions. It depends on fewer variables, for a fixed number of external legs, and is invariant under field redefinition. For similar reasons, we consider static patch correlators for operators that are pushed to its null boundaries, namely the past and future horizons. In tortoise coordinate
\be
x  = \arctanh \ r,
\ee
the horizon is at $x\to \infty$, and the in and out fields are defined as
\be
\phi_{\rm out}(u,\hat r) = \phi(t,x,\hat r),\qquad t\sim x \to \infty, \quad u \equiv t-x,
\ee
\be
\phi_{\rm in}(v,\hat r) = \phi(t,x,\hat r),\qquad -t\sim x \to \infty, \quad  v \equiv  t+x.
\ee
Some earlier works that motivate working with such in-out observables in the static patch include \cite{Anninos,Yasha,Law}.

To discuss quasinormal spectrum, we will focus on
\be\label{Fourier}
\hat{C}(\omega,\theta) \equiv \int_{-\infty}^\infty du e^{i\omega u} \expect{[\phi_{\rm out}(u,\hat r),\phi_{\rm in}(0,\hat z)]},
\ee
where $\theta$ is the angle between the two versors $\hat{r}$ and $\hat{z}$. In a free theory (\ref{Fourier}) is the Fourier transform of the retarded Green's function between two points approaching the past and future horizons. Note that there is no translation symmetry $x\to x+ \ep$. Therefore, even the two-point correlator is nontrivial. In a free theory, the above commutator is independent of the choice of state, but encodes the information about propagation of waves on the curved background. In an interacting theory, $\hat{C}(\omega,\theta)$ depends on the state. For instance, the Hartle-Hawking state is a thermal state in the static patch, and $\hat{C}(\omega,\theta)$ carries information about the interactions with the thermal excitations. 
\subsection{The free case}
\label{sec:free}
The exact dS$_{d+1}$ correlation function of a free scalar field is \cite{Spradlin}
\be\label{Gf}
G_f(\xi,\Delta) \equiv \expect{\phi(X) \phi(Y)}= \frac{\Gamma(\Delta) \Gamma(d-\Delta)}{(4\pi)^{\frac{d+1}{2}}\Gamma(\frac{d+1}{2})}
\hyp\left(\Delta ,d-\Delta;\frac{d+1}{2}; 1-\frac{1}{\xi}\right)
\ee
where $\Delta = \frac{d}{2} +i\sqrt{m^2 - \frac{d^2}{4}}$, and $X$ and $Y$ are the coordinates of a $d+2$-Minkowski spacetime that embeds dS$_{d+1}$ as the hyperboloid with $X\cdot X = Y\cdot Y = 1$. In terms of this Minkowski inner product
\be
\xi = \frac{2}{1- X\cdot Y}.
\ee
When $X$ and $Y$ are timelike separated $\xi<0$, and we are on the branch-cut of $G_f$. The two choices of the branch correspond to the two different ordering of the fields. The commutator is the discontinuity across this cut:
\be
[\phi(X),\phi(Y)] =-\Theta(-\xi) \frac{2\pi i}{(4\pi)^{\frac{d+1}{2}}\Gamma(\frac{3-d}{2})}(-\xi)^{\frac{d-1}{2}}
\left(1-\frac{1}{\xi}\right)^{\frac{1-d}{2}}\hyp\left(1-\Delta,1-d+\Delta,\frac{3-d}{2},\frac{1}{\xi}\right).
\ee
For the particular in-out configuration of the last section, 
\be\label{xi}
\xi = \frac{2}{1- \cos\theta-2 e^{u}},
\ee
and hence $\xi<0$ corresponds to $2e^u>1- \cos\theta$. After changing the integration variable in \eqref{Fourier} to $U= e^u - \frac{1-\cos\theta}{2}$, we find
\be\label{Cfree}
\hat{C}(\omega,\theta) = \frac{-2\pi i}{(4\pi)^{\frac{d+1}{2}}\Gamma(\frac{3-d}{2})}
\int_0^\infty dU \left(U+\frac{1-\cos\theta}{2}\right)^{i\omega-1} U^{\frac{1-d}{2}} (1+ U)^{\frac{1-d}{2}}
\hyp\left(1-\Delta,1-d+\Delta;\frac{3-d}{2};-U\right).
\ee
As a function of $\omega$, this integral is analytic in the upper half complex $\omega$ plane unless $\theta = 0$. Its singularities arise from the $U\to \infty$ limit of the integral. Indeed, the asymptotic expansion of the hypergeometric function is:
\be
\hyp(a,b;c;-x) \sim \frac{\Gamma(b-a)\Gamma(c)}{\Gamma(b)\Gamma(c-a)}x^{-a}+\frac{\Gamma(a-b)\Gamma(c)}{\Gamma(a)\Gamma(c-b)}x^{-b} \quad \text{as} \ x\to+\infty
\ee
Therefore the poles are located at
\be
\label{qnms}\begin{split}
& i\omega = \Delta + n,\\
& i\omega = d-\Delta + n,\qquad n\in \{0,1,2,\cdots\}.
\end{split}\ee
The explicit answer for \eqref{Cfree} is given in appendix \ref{app:explicitfree}.

The frequencies \eqref{qnms} are precisely the quasinormal mode spectrum of a scalar field on de Sitter \cite{Brady}. In dS, the quasinormal modes are defined as eigensolutions of the linear field equations with outgoing boundary condition at the cosmological horizon and regularity at the origin. They are called {\em quasinormal} because with these boundary conditions the linear differential operator is not Hermitian. Quasinormal modes characterize the decay of perturbations since their frequencies appear as the poles (in the lower half-plane) of the retarded Green's function. By the Cauchy theorem the response to a source $J$ at $t=0$ is \cite{Konoplya}
\be
\phi(t,\r) = \int d^3\r' \int_{-\infty}^{+\infty} \frac{d\omega}{2\pi}e^{-i\omega t} G_R(\omega,\r;\r')J(\r') = \sum_{{\rm QNMs}} a_n e^{-i\omega_n t} + \text{tail}.
\ee
The ``tail'' corresponds to non-exponential contributions (often power-law decays that eventually dominate) and comes from additional singularities of the Green's function. dS Green's function is special in that there are no other singularities except for the QNM poles, and hence there is no tail. We see that $\hat{C}(\omega,\theta)$ has the same analytic structure. 
\subsection{Spectrum of interacting theories}
The description of the quasinormal modes as the eigensolutions of the wave equation with particular boundary conditions is specific to free fields. However, we can still wonder about the analytic structure of the Green's function, or $\hat{C}(\omega,\theta)$, in an interacting theory. The poles would still describe the exponential decay of perturbations and hence they are the most natural generalization of the quasinormal spectrum to the interacting case. However, one might expect a more complicated singularity structure (such as branch-points) to appear at the interacting level. We can use the K\"all\'en-Lehmann representation for the 2-point correlators \cite{Hogervorst,DiPietro}
\be\label{XY}
\expect{\phi(X) \phi(Y)} = \int_{\frac{d}{2}-i \infty}^{\frac{d}{2}+ i\infty} \frac{d\Delta}{2\pi i}\rho(\Delta)
G_f(\xi;\Delta)+\cdots,
\ee
to evaluate $\hat{C}(\omega,\theta)$. In the above expression, the integral runs over the unitary representations of dS called the principal series. This is a redundant (but useful) presentation, since $\Delta\sim d-\Delta$ and $\rho(\Delta) =\rho(d-\Delta)$. By analytic continuation to $d+1$ sphere, one can show that $\rho^*(\Delta)= \rho(\Delta^*)$. Perturbative computations suggest that $\rho(\Delta)$ is generically meromorphic even in an interacting theory. Dots account for other unitary representations, such as the complementary series. In perturbative examples, those will form a discrete sum of $G_f(\xi;\Delta)$ with real $\Delta$.

We would like to connect the analytic structure of $\rho(\Delta)$ to that of $\hat{C}(\omega,\theta)$, by a contour deformation from the path $\Delta\in \frac{d}{2}+i\mathbb{R}$ closing it to the right half plane in \eqref{XY}. For this purpose it is useful to use the identity
\be
G_f(\xi;\Delta) = \frac{g(\Delta) \psi_\Delta(\xi)+(\Delta \to d-\Delta)}{2}
\ee
where
\be\label{gpsi}
g(\Delta) = \frac{\Gamma(\frac{d}{2}-\Delta)\Gamma(\Delta)}{2^{2\Delta+1}\pi^{d/2+1}},
\qquad
\psi_\Delta(\xi) = \xi^\Delta \hyp\left(\Delta,\Delta-\frac{d-1}{2},2\Delta - d+1,\xi\right).
\ee
Using the symmetry $\rho(\Delta) = \rho(d-\Delta)$, we can write \eqref{XY} as
\be\label{XY2}
\expect{\phi(X) \phi(Y)} = \int_{\frac{d}{2}-i \infty}^{\frac{d}{2}+ i\infty} \frac{d\Delta}{2\pi i}\tilde\rho(\Delta)
\psi_\Delta(\xi),
\ee
where
\be
\tilde \rho(\Delta) \equiv \rho(\Delta)g(\Delta).
\ee
In appendix \ref{app:contour}, we show that the contour in \eqref{XY2} can be deformed to the right if $ \rho$ decays faster than a $d$-dependent power of $\Delta$.\footnote{This is the same contour deformation performed in \cite{Hogervorst}, but instead of $\xi\to 0^+$ we are interested in the timelike configuration $\xi<0$, where there is no obvious exponential decay coming from $\psi_\Delta(\xi)$.} Assuming also that $\tilde\rho(\Delta)$ is meromorphic, we find
\be\label{sum}
\expect{\phi(X) \phi(Y)} = -\sum_i {\rm Res}_{\Delta_i}(\tilde \rho) \psi_{\Delta_i}(\xi).
\ee
Now the discontinuity of $\psi_\Delta$ in the timelike region comes entirely from the factor $\xi^\Delta$ in \eqref{gpsi}. The anti-symmetrized correlator $C(\xi,\theta)= \expect{[\phi(X),\phi(Y)]}$ has support only in this region:
\be
C(\xi,\theta) =-\Theta(-\xi) \sum_i 2i\sin(\pi\Delta) {\rm Res}_{\Delta_i}(\tilde \rho)
(-\xi)^{\Delta_i} \hyp\left(\Delta_i,\Delta_i-\frac{d-1}{2},2\Delta_i - d+1,\xi\right),
\ee
where $\Theta$ is the Heaviside function. As an example, we fix $\theta = \pi$ in \eqref{xi} and Fourier transform with respect to $u$ to find $\hat{C}(\omega,\pi)$ (except for $\theta =0$, at which $\phi_{\rm out}$ is in the future lightcone of $\phi_{\rm in}$, the singularity structure of $\hat{C}(\omega,\theta)$ is the same as $\hat{C}(\omega,\pi)$). A single term in the sum \eqref{sum}, after the change of variable $V=-\xi = 1/(e^u-1)$, gives
\be
\int_0^\infty dV (1+V)^{i\omega -1} V^{\Delta_i-i\omega-1 } \hyp\left(\Delta_i,\Delta_i-\frac{d-1}{2},2\Delta_i - d+1,-V\right).
\ee
As a function of $\omega$, the singularities of $\hat{C}(\omega,\pi)$ arise from the $V\to 0$ region of integration. These are poles located at
\be
i\omega_n = \Delta_i + n, \qquad n\in \{0,1,2,\cdots\}.
\ee
Given that $\Re(\Delta_i)\geq d/2$, these poles are all in the lower half-plane as expected from causality. Hence, if $\tilde \rho$ is meromorphic, for every pole $\Delta_i$, we have the full family of the poles corresponding to the QNMs spectrum of a fictitious field with dimension $\Delta_i$ (thought not $d-\Delta_i$). The real analyticity of $\tilde\rho$ implies the existence of another family with associated to $\Delta_i^*$. The overtones (frequencies with $n \ge 1$) can be thought of as arising from the descendants of a conformal primary operator with dimension $\Delta_i$. Similarly, for every complementary series representations that appears in \eqref{XY}, there will be the full set of associated QNMs. 
\subsection{A continuous spectrum}
\label{sec:counter}
In \cite{DiPietro}, the in-in perturbation theory was analytically continued to one in Euclidean AdS. This automatically results in a meromorphic $\rho(\Delta)$, and a discrete sum over complementary series representations. Hence, if there is a counter-example, it must be non-perturbative. It is well known that for light interacting scalar fields in dS, perturbation theory fails. Originally, this problem was solved by switching to a stochastic description \cite{Starobinsky,SY}. This allows a reliable non-perturbative computation of the relaxation exponents as the eigenvalues of the Fokker-Planck equation
\be
\d_tp(t,\vphi) = \frac{1}{8\pi}\d_\vphi^2p(t,\vphi) + \frac{1}{3} \d_\vphi(V'(\vphi)p(t,\vphi))+\cdots
\ee
where $V(\vphi)$ is the scalar potential and dots include corrections that can be computed perturbatively (in slow-roll parameters such as $V'^2/H^2V, V''/H^2$) \cite{Gorbenko,Markov,Cohen}. The eigenvalue problem can be mapped to a Schr\"odinger equation with an effective potential \cite{SY}
\be
V_{\rm eff}(\vphi) = \frac{1}{2}(v'^2(\vphi) - v''(\vphi)),\qquad v(\vphi) \equiv \frac{4\pi^2}{3}V(\vphi).
\ee
If we take $v(\vphi) \propto \vphi \tanh \vphi$, the effective potential approaches a constant as $\vphi\to \pm \infty$ and hence the spectrum of decay exponents will have a continuum above a gap. It follows that the corresponding $\rho(\Delta)$ is not meromorphic in this example. For an application of this model see \cite{George_pbh}.
\subsection{Discrete spectrum in Rindler space}
\label{sec:rindler}
As a simpler example, we consider the retarded function in the Rindler wedge of $2d$ Minkowski spacetime. We will see that generic interacting theories will have a discrete spectrum of singularities in terms of the Rindler frequency. As before, we define the in and out fields by taking the fields to the past and future Rindler horizons. The metric in the left Rindler wedge can be written as
\be
ds^2 =4 e^{-2x}(-dt^2 + dx^2).
\ee
The Minkowski coordinates are related to these by
\be
X^0 = 2 e^{-x} \sinh t,\qquad X^1 =- 2 e^{-x} \cosh t.
\ee
The past and future horizons correspond to $x\to \infty$ and $t\to \mp \infty$:
\be\begin{split}
X^0 =& - X^1 = e^{t-x}\equiv e^{u},\qquad \text{future}\\[10pt]
X^0 = & X^1 = -e^{-t-x} \equiv e^{-v}.\qquad \text{past.}
\end{split}
\ee
We define $\phi_{\rm in}(v)$ and $\phi_{\rm out}(u)$ as the fields pushed to the horizons and study
\be\label{commute}
C(u) \equiv \expect{[\phi_{\rm out}(u),\phi_{\rm in}(0)]}
\ee
Note that the expectation values in \eqref{commute} could be thought of as being computed in the thermal state obtained after tracing out the complement of the Rindler wedge, or as Minkowski correlators. We are interested in the analytic structure of the Fourier transform of $C(u)$:\footnote{From the perspective of Minkowski coordinates, this transform is similar to going to rapidity space and $\hat{C}(\omega)$ resembles the celestial amplitudes. For two recent works on the analytic properties of such amplitudes see \cite{Duary,Kapec}.}
\be
\hat{C}(\omega) = \int_{-\infty}^{+\infty}
  du e^{i\omega u} C(u).
\ee
Let us compute this in a theory with mass gap $m$ and a density of states $\rho(s)\geq 0$. The K\"all\'en-Lehmann representation of the Minkowski Feynman correlator is
\be
\expect{T\{\phi(X)\phi(0)\}} = \int_{\mathbb{R}^2} d^2 k \int_0^\infty  \frac{ds}{2\pi} \frac{\rho(s)}{k^2+s+i\ep} e^{ik\cdot X}.
\ee
The commutator \eqref{commute} is the difference between the time-ordered and anti-time-ordered correlators, which implies (after switching to $k_\pm = k^0 \pm k^1$)
\be
\hat{C}(\omega) =\frac{1}{2} \int_{-\infty}^\infty du \int dk_+ dk_- \int_{0}^\infty ds \rho(s)\delta(k_+ k_- - s)  e^{i\omega u-ik_- e^u - i k_+}.
\ee
It is straightforward to perform $k_+$, then $u$ and then $k_-$ integrals. For instance, the $u$ integral can be rewritten in terms of $U\equiv e^u$ as
\be
\int_0^\infty dU U^{i\omega -1} e^{-i k_- U} = \Gamma(i\omega) (ik_-)^{-i\omega}.
\ee
The final result is
\be\label{C}
\hat{C}(\omega) = \frac{1}{2\pi} \sinh(\pi \omega) \Gamma^2(i \omega) \int_{0}^\infty ds \rho(s)  s^{-i\omega}.
\ee
In a free theory $\rho(s)= \delta(s- m^2)$ and we see that $\hat{C}(\omega)$ is meromorphic. In this case it has poles in the upper half plane. Given that the fields are always timelike separated for any $u$, there is no reason for the Fourier transform to be analytic in the upper-half-plane.\footnote{In fact, the function decays fast enough in the upper-half $\omega$ plane that the inverse Fourier transform for both signs of $u$ can be computed by closing the integration contour on that side.}

In the interacting case, one expects additional singularities to arise in \eqref{C}. Now $\rho(s)$ is in general a continuous function. For instance, it accounts for the multi-particle states that mix with the single particle states via interactions. However, if the theory is gapped, i.e. $\rho(s<m^2) =0$, then the additional singularities in \eqref{C} can only arise from the $s\to \infty$ limit of the integral. Generically, $\rho(s)$ behaves as a power (or a combination of powers) of $s$. This would again lead to a discrete set of poles in $\omega$. 

In general the same thing does not apply to dS: if the spectral function is not meromorphic, then in the deformation of the contour in the integral (\ref{XY}) one must take into account the contribution coming from branchcuts, that result in integral of some discontinuities. Generically, this cannot be represented as a discrete sum over pure exponentials.
\section{Boundary effective field theory}
\label{sec:EFT}
A useful setup for studying low-frequency perturbations in a black hole geometry is the worldline EFT \cite{Porto}. The idea is that for long wavelength perturbations the neighborhood of a black hole can be replaced by a worldline on a weakly curved geometry and with interactions localized on it. This is an open system because waves can fall through the horizon. The effect can be taken into account by introducing additional operators on the worldline that represent near horizon degrees of freedom and are short-lived \cite{Nicolis,Porto}. One might expect a local dissipative EFT, formulated on the Schwinger-Keldysh (SK) contour, to arise after integrating out these operators \cite{Liu}. 

In this section, we ask if a similar idea can be applied to the low-frequency in-out observables in dS static patch. The analog of the black hole region is now the region $0<x\sim 1$. Since the size of $S^{d-1}$ factor reaches an $\O(1)$ constant as $x\to \infty$, the appropriate EFT would be a collection of KK modes propagating on a half-line, and with interactions localized at the boundary. The EFT is dissipative because we are interested in a description of modes with $\omega\ll 1$ in a system at finite temperature $T=\frac{1}{2\pi}$.

\subsection{Free theory}
To motivate the idea, consider a massive free field and decompose it in terms of spherical harmonics
\be
\phi(t,\r) =  \sum_{l,m} \phi_{lm}(t,x) Y_{lm}(\hat r).
\ee
In a free theory, different $\phi_{lm}$ decouple. The action for the s-wave perturbations is
\be\label{S00}
S_{00} = \frac{1}{2} \int dt \int_0^\infty dx \tanh^2 x \left[\dot\phi_{00}^2- {\phi_{00}'}^2- \frac{m^2}{\cosh^2 x} \phi_{00}^2\right].
\ee
We expand $\phi_{00}$ in radial eigenmodes
\be
\phi_{00}(t,x) = \int_0^\infty \frac{d\omega}{2\pi} \phi_\omega(t) f_\omega(x),
\ee
where asymptotically
\be
f_\omega(x \to \infty) = e^{i\omega x + i\alpha} + e^{-i\omega x - i\alpha},
\ee
and the phase shift is given by\footnote{For generic $l$ the mode function is
\be
\phi_{\omega l}(t,x) \propto e^{-i\omega t} \tanh^l x (\cosh x)^{-i\omega} \hyp\left(\frac{l+\Delta_++i\omega}{2},\frac{l+\Delta_-+i\omega}{2},\frac{d}{2}+l,\tanh^2 x\right).
\ee
For an application of the dS phase shift and its analogs on black hole backgrounds to the computation of the free field partition function see \cite{Anninos,Law}.}
\be\label{alpha}
e^{2i\alpha} = \frac{2^{-2 i\omega} \Gamma(i\omega) \Gamma(\frac{1}{2} (\Delta_+ -i \omega))
\Gamma(\frac{1}{2} (\Delta_- -i \omega))}
{\Gamma(-i\omega) \Gamma(\frac{1}{2} (\Delta_+ +i \omega))
\Gamma(\frac{1}{2} (\Delta_- +i \omega))}.
\ee
Now every $\phi_\omega(t)$ is a harmonic oscillator with frequency $\omega$ and the Hartle-Hawking state corresponds to a thermal state with $T=1/2\pi$. The ladder operators, defined via $\phi_\omega(t) = \frac{1}{\sqrt{2\omega}}\left(a_\omega e^{-i\omega t} + a_\omega^\dagger e^{i\omega t} \right)$, satisfy
\be
\expect{a_\omega^\dagger a_{\omega'}} = 2\pi \delta(\omega -\omega') \frac{1}{e^{2\pi\omega} -1}.
\ee
The in-out correlator is given by
\be\label{contour}
\expect{\phi_{00}^{\rm out}(u) \phi_{00}^{\rm in}(0)} = 
\Cint \frac{d\omega}{2\pi } e^{-i\omega u} \frac{e^{2i\alpha}}{2\omega \left(1-e^{-2\pi \omega}\right)},
\ee
where the integration contour runs from $-\infty$ to $+\infty$ and below the double pole at $0$. Naively, one gets a singular integral that goes through $0$. This singularity is an artifact of the exchange of the limit $x\to \infty$ and the integral. The contour deformation is the one that reproduces the correct result. 

What suggests the possibility of an effective worldline theory that matches this result at low $\omega$ is that at large $x$ the action \eqref{S00} reduces to that of a massless $2d$ field on a half-line. We can couple this to a boundary by writing (we drop the $00$ index to avoid clutter)
\be\label{EFT}
S_{\rm BEFT} =\int dt \int_0^\infty dx \ \left[\frac{1}{2}\left(\dot\phi^2 -{\phi'}^2\right) -\delta(x) \sum_n c_{n} \left(\d_t^n\phi\right)^2\right].
\ee
We now show that with an appropriate choice of $c_n$ this BEFT matches the expansion around $\omega =0$ of the full result. It is enough to match the low-$\omega$ expansion of the phase-shifts on the two sides. The equation for the mode functions in the EFT reads
\be
f_\omega''(x) + \omega^2 f_\omega(x) = \delta(x) f_\omega(0) \sum_{n=0}^\infty c_n \omega^{2n}.
\ee
The correctly normalized solution is $f_\omega(x)= 2 \cos(\omega x + \alpha)$ where
\be\label{tan}
\omega \tan \alpha=i \omega \frac{1-e^{2i\alpha}}{1+e^{2i\alpha }} = -\sum_{n=0}^\infty c_n \omega^{2n}.
\ee
A unique choice of $\{c_n\}$ exists because the phase shift given in \eqref{alpha} is real for real $\omega$ and satisfies
\be
e^{2i\alpha(0)} = -1,\qquad e^{2i\alpha(-\omega)} = e^{-2i \alpha(\omega)}.
\ee
\subsection{Interacting theory}
\label{sec:intBEFT}
It is natural to ask if the in-out correlators of an interacting theory can also be matched to a BEFT, and what constraints unitarity imposes on the EFT parameters. Intuitively, this would be the case if interactions turn off fast enough for $x\gg 1$. Then for the low-frequency observables the interaction region $x\sim 1$ can be collapsed into a point at $x=0$. 

It would be interesting to find a classification of theories for which this is possible. We did not succeed so far, but we do not expect the set to be empty either. It plausibly includes super-renormalizable theories because of the large blue-shift toward the horizon. The example of last section supports this expectation: We can think of the mass term as an interaction added to the free massless theory. As we showed explicitly, the correlators of this theory match with a massless free theory coupled to a boundary.

Given $\rho(\Delta)$, a way to check the rapid fall-off of interactions is to inspect the projection of \eqref{XY} into the space of spherical harmonics
\be\label{in-out}
\expect{[\phi_{lm}^{\rm out}(u), \phi_{lm}^{\rm in}(0)]} = 
\int_{\frac{d}{2}-i \infty}^{\frac{d}{2}+ i\infty} \frac{d\Delta}{2\pi i}\rho(\Delta)
C_f(u,l;\Delta).
\ee
If the interactions turn off at large $x$, we would expect the LHS to be universal for $u\ll -1$ because an in-state that travels all the way to $x\sim 1$ region and reflects back arrives to the future horizon at $u$ not much less than $-1$. We note that regardless of $\Delta$, all free commutators $C_f(u,l;\Delta)$ approach a universal form in the limit $u\to -\infty$:
\be
C_f(u,l;\Delta) \to -\frac{i}{2} + \O(e^u),\qquad u\ll -\log|\Delta|.
\ee
Therefore, if the integration over $\Delta$ converges with a negative power of $|\Delta|$, the interacting commutator will approach the same universal behavior exponentially in $u$.\footnote{One advantage of S-matrix is its invariance under field redefinitions. This does not hold for static-patch in-out correlators since, from a global perspective, they are just a particular configuration of local correlators. For instance, changing $\phi\to \phi + \phi^2$ in a free theory results in $\rho(\Delta)=\rho_\phi(\Delta)+\rho_{\phi^2}(\Delta)$ that given $\rho_{\phi^2}(\Delta)$ (see \cite{Epstein,Hogervorst}) does not satisfy the above convergence requirement.} Since nontrivial interactions would ultimately leave their imprint in this observable, if the latter is universal for $u\ll -1$, we expect a boundary description to emerge at low frequency. Assuming this, we proceed to show that unitarity (i.e. $\rho(\Delta)>0$) constrains this BEFT. 

First, we write \eqref{in-out} as
\be\label{Comint}
\expect{[\phi^{\rm out}_{lm}(u), \phi^{\rm in}_{lm}(0)]} = 
\Cint \frac{d\omega}{4\pi \omega}e^{-i\omega u} \mathcal{S}(\omega),
\ee
where the contour is the same as in \eqref{contour} and 
\be\label{Sint}
\S(\omega) \equiv \int_{\frac{d}{2}-i \infty}^{\frac{d}{2}+ i\infty} \frac{d\Delta}{2\pi i}\rho(\Delta) e^{2i\alpha_\Delta(\omega,l)}.
\ee
Above, to exchange the order of the $\omega$ and $\Delta$ integrals, we assumed a sufficiently fast convergence of the integral over $\Delta$. More generally, one might need to perform a number of subtractions before being able to do so. Depending on $l$, the free phase shifts $e^{2i\alpha_\Delta(\omega,l)}$ approach $1$ or $-1$ as $\omega\to 0$. In an interacting theory, $\rho(\Delta)$ is supported on more than one $\Delta$. Therefore, there has to be nonzero absorption:
\be\label{norm}
|\S(\omega)| < |\S(0)|,\qquad \text{interacting theory.}
\ee
Let us focus on the $l=0$, $d=3$ case and try to match the low $\omega$ behavior of $\S(\omega)$ with a BEFT. In an interacting theory, the boundary action will include interactions and not just quadratic operators. Moreover, this effective theory includes the novel features of a finite temperature EFT, namely dissipative terms that have to be formulated on the SK contour. Absorption is caused not only by the production of the low-energy effective degrees of freedom, but also via interaction with the thermal bath. 

It is believed that when the environment degrees of freedom have a short correlation time $\tau$, the long wavelength ``hydro description'' admits a local expansion controlled by $\tau$ \cite{Liu}. For us $\tau$ is related to the quasinormal decay time. We have seen that the spectrum is typically discrete and $\tau \sim H^{-1} (= 1)$. This implies the existence of a local expansion for $\omega \ll 1$
\be\label{Sexpand}
\S(\omega) = -\sum_{n=0}^\infty \frac{1}{n!} s_n (-i\omega)^n
\ee
(the overall minus is for later convenience). Since this is a 2-point correlator, the coefficients $s_n$ could be matched with a quadratic action. However, in addition to \eqref{EFT}, we have to include the dissipative terms to the action (dropping as before the $00$ index of the $s$-wave)
\be
\label{eq:dissipative}
S^{(2)}_{\rm diss.} = 
-\int dt \int_0^\infty dx \delta(x) \sum_{n=0}^\infty b_{n} \d_t^n\phi_- \d_t^{n+1}\phi_+,
\ee
where in terms of the fields on the forward and backward part of the SK contour (respectively $\phi_R$ and $\phi_L$)
\be
\phi_+ = \frac{\phi_L+ \phi_R}{2}, \qquad \phi_-= \phi_R-\phi_L.
\ee
Following the SK perturbation theory (see Appendix \ref{app:SK}), we compute the leading dissipative correction 
\be
\S^{\rm EFT}(\omega) = e^{2i \alpha} \left[1-\frac{2 b_0}{c_0^2} \omega^2 +\O(\omega^3)\right],
\ee
where $\alpha$ is fixed by the conservative coefficients as in \eqref{tan}. Note that $\S^{\rm EFT}(0) = -1$. This means that matching is possible only if 
\be
\int_{\frac{d}{2}-i \infty}^{\frac{d}{2}+ i\infty} \frac{d\Delta}{2\pi i}\rho(\Delta) = 1.
\ee
Furthermore, in the presence of interactions \eqref{norm} implies a postive friction term 
\be
b_0>0.
\ee
It is of course natural to have a friction rather than an anti-friction in an open system.

There is an alternative way to find this bound and new ones on the higher derivative conservative terms using the moments constraints. First, we expand the free phase shifts \eqref{alpha}:
\be
e^{2i\alpha_\Delta(\omega,l=0)} =- \sum_{n=0}^\infty \frac{1}{n!} s_n(\Delta) (-i\omega)^n.
\ee
For $m^2 >2$ (i.e. above the conformal mass), we have 
\be\label{sns}
s_n(\Delta) >0,\qquad s_1(\Delta) = [s_0(\Delta)]^2,\qquad s_2(\Delta) \geq \frac{2}{3} [s_0(\Delta)]^3.
\ee
Substituting the expansion in \eqref{Sint},
\begin{equation}
\S(\omega) = -\sum_{n=0}^\infty \left[\int_{\frac{d}{2}-i\infty}^{\frac{d}{2}+i\infty} \frac{d\Delta}{2\pi i} \, 
\rho(\Delta) s_n(\Delta)\right] (-i\omega)^n,
\end{equation}
we can write the coefficients as averages of over the following distribution:
\begin{align}
    \int_{\frac{d}{2}-i \infty}^{\frac{d}{2}+ i\infty} \frac{d\Delta}{2\pi i} \, \rho(\Delta)s_n(\Delta) = \int_0^\infty d s_0 \underbrace{\frac{d\Delta}{d s_0}\frac{\rho\left(\Delta(\alpha_n)\right)}{2\pi i}}_{\equiv \hat{\rho}(s_0)>0} s_n(s_0) \equiv \langle s_n\rangle.
\end{align}
On the other hand, these coefficients must match the EFT result (see Appendix \ref{app:SK}):
\begin{align}\label{SEFT0}
-\S^{\rm EFT}(\omega) = 1+\frac{2}{-c_0}(-i\omega)+\frac{2(b_0+1)}{c_0^2}(-i\omega)^2+\frac{2(1+2b_0+b_0^2+c_0c_1)}{-c_0^3}(-i\omega)^3+\mathcal{O}\left(\omega^4\right),
\end{align}
where $c_0 <0$ if $\rho$ has support only for $m^2>2$. The second relation in \eqref{sns} and the fact that $\expect{s_0^2}\ge \expect{s_0}^2$ implies
\begin{align}
    \frac{1+b_0}{c_0^2} \ge \frac{1}{c_0^2} \quad \Leftrightarrow \quad \boxed{b_0 \ge 0}.
\end{align}
The third relation in \eqref{sns} and $\langle s_0\rangle\left\langle s_0^3\right\rangle \ge \left\langle s_0^2\right\rangle^2$ implies
\begin{align}
    \boxed{c_1 \le \frac{(1+b_0)^2}{-3c_0}}.
\end{align}
So far we did not include boundary interactions. Loops from those would lead to additional analytic contributions to \eqref{SEFT0} starting from $\O(\omega)$. Therefore, they are degenerate with the $b_n$ and $c_n$ parameters, and our bounds should be interpreted as unambiguous constraints on the expansion coefficients of $\S(\omega)$ rather than coefficients in the Lagrangian. The scheme dependence of the Lagrangian is, of course, a general fact. 
\section{Conclusions}
\label{sec:con}
In this paper, we worked with in-out correlators in dS static patch to give a generalization of the quasinormal spectrum in interacting theories. We related this spectrum to the spectrum of dimensions that appear in the K\"al\'en-Lehmann spectral density $\rho(\Delta)$ of the 2-point correlators. We have seen that the quasinormal spectrum is generically (though not always) discrete, and the analytic structure of the correlators is similar to the celestial amplitudes. 

The downside of our approach is that the horizon correlators are not well-defined when gravity is dynamical. In that context, it seems more natural to define the operator insertions with respect to the worldline of a dS observer. On the other hand, the generalization of the quasinormal spectrum to the interacting case might be relevant in computing the corrections to de Sitter entropy as thoroughly discussed in \cite{Anninos}.

We have also showed that at low frequencies, the in-out correlators can, under certain conditions, be matched to a finite temperature boundary EFT. This BEFT can be thought of as the hydrodynamic limit of the problem. It might be useful in understanding the structure of thermal EFTs and dissipative hydrodynamics. In particular, it would be interesting to find similar positivity bounds in them. 

Finally, it would be interesting to explore the consequences of unitarity for the cosmological correlation functions, and to find potentially observable consistency conditions. See \cite{Goodhew} for an earlier work in this direction. 

\section*{Acknowledgments}
We thank Dionysios Anninos, Lorenzo Di Pietro, Sergei Dubovsky, Victor Gorbenko, Shota Komatsu, and Kamran Salehi Vaziri for useful discussions.
\appendix
\section{Free commutator}
\label{app:explicitfree}
Starting from eq. (\ref{Cfree}) and setting $\Delta=\frac{d}{2}+i\nu$:
\be
    \hat{C}(\omega,\theta) = \frac{2\pi i}{(4\pi)^\frac{d+1}{2}\Gamma\left(\frac{3-d}{2}\right)}\mathcal{M}\left[v\mapsto\left(v+\frac{1-\cos\theta}{2}\right)^{i\omega-1}{}_2F_1\left(\frac{1}{2}-i\nu,\frac{1}{2}+i\nu;\frac{3-d}{2};-v\right)\right]\left(\frac{3-d}{2}\right),
\ee
where $\mathcal{M}[f](s)=\int_0^\infty x^{s-1}f(x) dx$ is the Mellin transform that can be analytically computed.
\begin{align}
\label{eq:thetacommutator}
    & \hat{C}(\omega,\theta) = \frac{i\Gamma\left(\frac{d+1}{2}-i\omega\right)\Gamma\left(\frac{1-d}{2}+i\omega\right)}{2^d \pi^\frac{d-1}{2}\Gamma\left(1-i\omega\right)}\times\\
    &\times\left[\frac{\Gamma\left(\frac{d}{2}+ i\nu-i\omega\right)\Gamma\left(\frac{d}{2}- i\nu-i\omega\right)}{\Gamma\left(\frac{1}{2}+ i\nu\right)\Gamma\left(\frac{1}{2}- i\nu\right)}{}_2\tilde{F}_1\left(\frac{d}{2}+i\nu-i\omega,\frac{d}{2}-i\nu-i\omega,\frac{d+1}{2}-i\omega,\sin^2 \frac{\theta}{2}\right)+\right.\\
    &\left.-\left(\sin^2 \frac{\theta}{2}\right)^{i\omega+\frac{1-d}{2}}{}_2\tilde{F}_1\left(\frac{1}{2}+i\nu,\frac{1}{2}-i\nu,\frac{3-d}{2}+i\omega,\sin^2 \frac{\theta}{2}\right)\right],
\end{align}
where ${}_2\tilde{F}_1$ is the regularized hypergeometric function that has no singularities.\comment{\footnote{The hypergeoemtric function is defined, for $|z|<1$ as:
\begin{align}
    {}_2F_1(a,b;c;z) = \sum_{n=0}^\infty \frac{(a)_n (b)_n}{(c)_n}\frac{z^n}{n!}
\end{align}
where $(q)_n = \frac{\Gamma(q+n)}{\Gamma(q)}=\prod_{k=0}^{n-1}(q+k)$ is the Pochhammer symbol, which has no poles but it has simple zeros for $q=0,-1,\dots,n-1$, $\forall n\ge 1$. Therefore the only pole in the $a,b,c$ variables is when $c=-n$, with $n\in\mathbb{N}$. However, the regularized hypergeometric function ${}_2\tilde{F}_1(a,b;c;z) =\frac{{}_2F_1(a,b;c;z)}{\Gamma(c)}$ has no poles, provided that $|z|<1$. For $z=1$ instead there could be poles because:
\begin{align}
\label{eq:hypergeometricat1}
    {}_2\tilde{F}_1(a,b;c;1)=\frac{\Gamma(c-a-b)}{\Gamma(c-a)\Gamma(c-b)}.
\end{align}}}
From the expression (\ref{eq:thetacommutator}) it is straightforward the evaluation for $\theta=0,\frac{\pi}{2},\pi$, where we can use that:
\begin{align}
    & {}_2\tilde{F}_1(a,b;c;0)=\frac{1}{\Gamma(c)}, \\
    & {}_2\tilde{F}_1\left(a,1-a;c;\frac{1}{2}\right)=\frac{\Gamma\left(\frac{c}{2}\right)\Gamma\left(\frac{c+1}{2}\right)}{\Gamma\left(c\right)\Gamma\left(\frac{c+a}{2}\right)\Gamma\left(\frac{c-a+1}{2}\right)},\\
    & {}_2\tilde{F}_1(a,b;c;1)=\frac{\Gamma(c-a-b)}{\Gamma(c-a)\Gamma(c-b)},
\end{align}
to show that:
\begin{align}
\label{eq:freecommutatoralign}
    & \hat{C}(\omega,0) = \frac{i\Gamma\left(\frac{1-d}{2}+i\omega\right)\Gamma\left(\frac{d}{2}+ i\nu-i\omega\right)\Gamma\left(\frac{d}{2}- i\nu-i\omega\right)}{2^d \pi^\frac{d-1}{2}\Gamma\left(1-i\omega\right)\Gamma\left(\frac{1}{2}+ i\nu\right)\Gamma\left(\frac{1}{2}- i\nu\right)},\\
    & \hat{C}\left(\omega,\frac{\pi}{2}\right) = \frac{i\Gamma\left[\frac{1}{2}\left(\frac{d}{2}+ i\nu-i\omega\right)\right]\Gamma\left[\frac{1}{2}\left(\frac{d}{2}-i\nu-i\omega\right)\right]}{2^{2(1+i\omega)} \pi^\frac{d}{2}\Gamma\left(1-i\omega\right)},\\
    & \hat{C}(\omega,\pi) = \frac{i\Gamma\left(\frac{d}{2}+ i\nu-i\omega\right)\Gamma\left(\frac{d}{2}- i\nu-i\omega\right)}{2^d \pi^\frac{d-1}{2}\Gamma\left(1-i\omega\right)\Gamma\left(\frac{d+1}{2}- i\omega\right)}.
\end{align}
The poles are those in eq. (\ref{qnms}). For $\theta=0$ there is an additional set of poles that extends to the upper half plane:
\be
\omega=i\left(n+\frac{1-d}{2}\right), \ n\in\mathbb{N}.
\ee
For $\theta \ne 0$ microcausality condition ensure the analyticity in the upper half plane. Indeed the condition $\left[\phi(X),\phi(Y)\right]=0$ when $(X-Y)^2 >0$ means that the in-out commutator vanishes for $u<u_0(\theta)=\log\left(\sin^2\frac{\theta}{2}\right)$.
If we anti-Fourier transform $\hat{C} (\omega,\theta)$ to have the commutator in the real space:
\begin{align}
    C(u,\theta)=\int_{-\infty}^{+\infty} \frac{d\omega}{2\pi} e^{-i\omega u}\hat{C}(\omega,\theta),
\end{align}
for $\theta=\pi$, $u_0(\pi)=0$, then:
\begin{align}
    C(u,\pi)=\int_{-\infty}^{+\infty} \frac{d\omega}{2\pi} e^{-i\omega u}\hat{C}(\omega,\pi)=0 \text{ for } u<0.
\end{align}
If $u<0$ we close the contour upwards and this means that $\hat{C}(\omega,\pi)$ has to be analytic in the upper half plane because of the residue theorem.

If $\theta \in (0,\pi)$:
\be
    C(u,\theta)=\int_\mathbb{R} \frac{d\omega}{2\pi} e^{-i\omega\left(u-u_0(\theta)\right)}e^{-i\omega u_0(\theta)}\hat{C}(\omega,\theta) = 0 \text{ for } u<u_0(\theta).
\ee
Again, for $u<u_0(\theta)$ we close the contour upwards and then the function $\omega\mapsto e^{-i\omega u_0(\theta)}\hat{C}(\omega,\theta)$ has to be analytic in the upper-half plane, but since $\omega\mapsto e^{-i\omega u_0(\theta)}$ is an entire function, therefore $\omega\mapsto \hat{C}(\omega,\theta)$ has to be analytic.

The only failure of this argument is for $\theta\to 0$, because $u_0(\theta) \to -\infty$ and so the microcausality condition is not applicable, since the commutator can be non-vanishing $\forall u\in\mathbb{R}$, that is why its Fourier transform $\hat{C}(\omega,\theta)$ can have poles in the upper-half plane, as indeed happens e.g. in the free case in eq. (\ref{eq:freecommutatoralign}).

The analytic structures of $\hat{C}(\omega,0)$ and $\hat{C}(\omega,\pi)$ for the free theory are reported in figure \ref{fig:poles}.

\begin{figure}[h]
\centering
  \includegraphics[scale=.6]{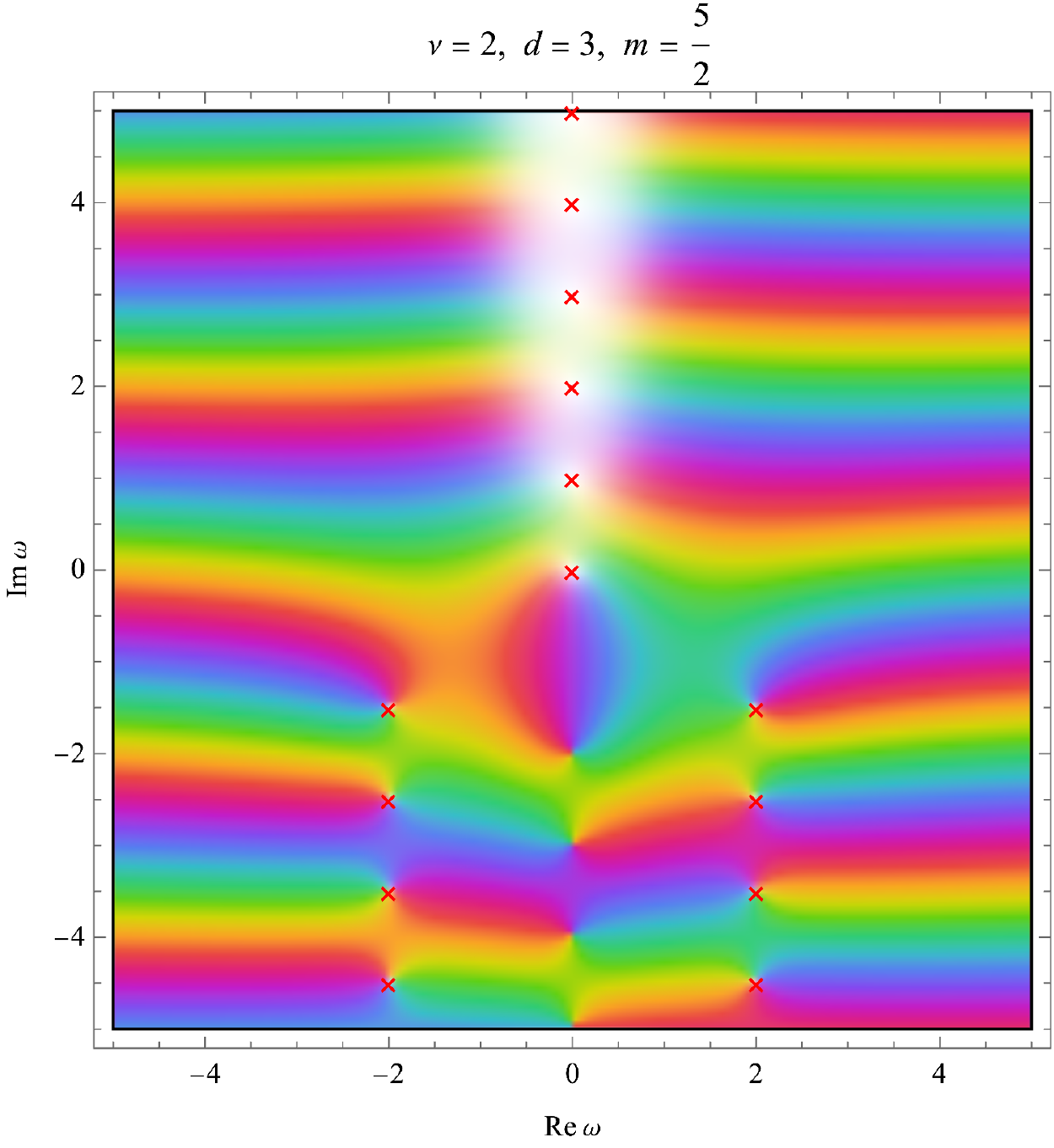}
  \includegraphics[scale=.6]{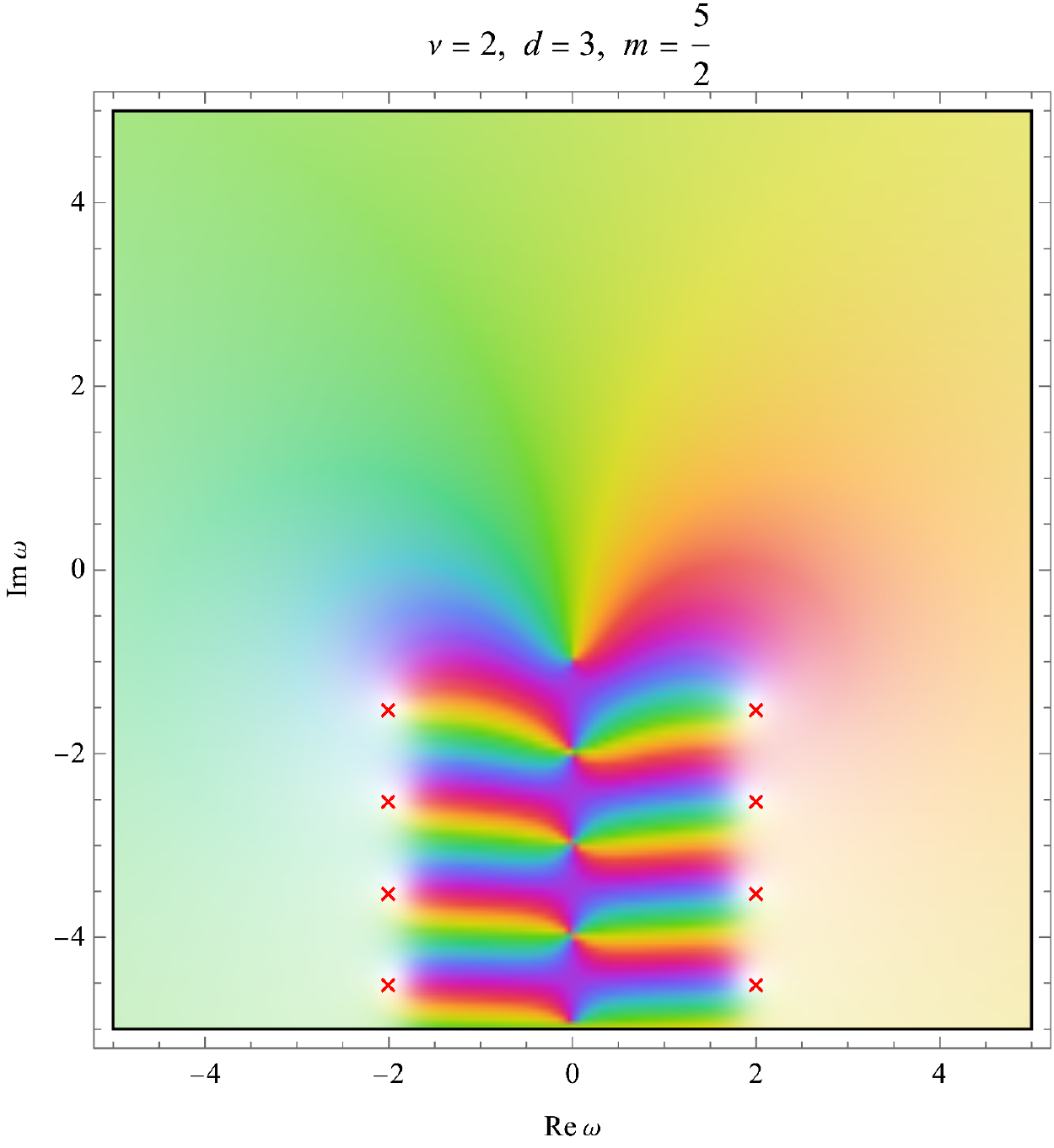}
  \caption{Complex plot of $\mathrm{Arg}\left[\hat{C}(\omega,\theta)\right]$ vs $\omega$. Left $\theta=0$ and right $\theta=\pi$. The red crosses highlight the poles.}
  \label{fig:poles}
\end{figure}
\section{Closure of the contour}
\label{app:contour}
We want to show under which conditions the integral in eq. (\ref{XY2}) can be computed via contour deformation to use the residue theorem. It is sufficient to prove that the integral on the arc at infinity vanishes. In terms of $\nu$ defined via $\Delta = \frac{d}{2}-i\nu$, the condition is
\begin{align}
\label{eq:limit}
    \left|\lim_{R\to\infty} R\int_0^\pi d\theta \ \rho\left(\frac{d}{2}-i\nu\right)g\left(\frac{d}{2}-i\nu\right)\psi_{\frac{d}{2}-i\nu}\left(\xi\right)\bigg|_{\nu=Re^{i\theta}}\right|=0.
\end{align}
We make a technical hypothesis on $\rho$ that is:
\be
\label{eq:hypothesis}
\left|\nu \rho\left(\frac{d}{2}-i\nu\right)\right| \lesssim \left|\frac{\Gamma\left(\frac{3}{2}-i\nu\right)}{\Gamma\left(\frac{d}{2}-i\nu\right)}\right| \quad \text{as} \quad |\nu|\to\infty,
\ee
otherwise one can see that the (\ref{eq:limit}) is not zero. This condition imposes $d$-dependent power-like decay on $\rho$. The perturbative examples in $d=2,3$ satisfy this requirement. 

The hypergeometric function can be approximated via the saddle point approximation of its integral representation:
\begin{align}
    &\left|{}_2F_1\left(\frac{d}{2}-i\nu,\frac{1}{2}-i\nu;1-2i\nu;\xi\right)\right|=\\
    &=\left|\frac{\Gamma(1-2i\nu)}{\Gamma^2\left(\frac{1}{2}-i\nu\right)}\int_0^1 dx \left[x(1-x)\right]^{-\frac{1}{2}}\left(1-\xi x\right)^{-\frac{d}{2}}\exp\left[-i\nu\log\frac{x(1-x)}{1-\xi x}\right]\right|\sim\\
    & \sim \left|a(\xi) \frac{\Gamma(1-2i\nu)}{\Gamma^2\left(\frac{1}{2}-i\nu\right)} \frac{1}{\sqrt{\nu}}\frac{1}{\left(2-\xi+2\sqrt{1-\xi}\right)^{-i\nu}}\right|
\end{align}
One finds that, for $|\nu|\to\infty$:
\begin{align}
    \left|g\left(\frac{d}{2}-i\nu\right)\psi_{\frac{d}{2}-i\nu}(\xi)\right| \sim A(\xi) \left|\frac{\Gamma\left(\frac{d}{2}-i\nu\right)}{\Gamma\left(\frac{1}{2}-i\nu\right)\sin(i\pi\nu)}\right||\nu|^{-\frac{1}{2}}e^{\pi\mathrm{Re}\nu} [c(\xi)]^{\mathrm{Im}\nu}
\end{align}
where $A(\xi)$ is a constant and:
\begin{align}
    c(\xi)=\frac{-\xi}{2-\xi+2\sqrt{1-\xi}} \in (0,1), \quad \text{for} \ \xi\in(-\infty,0)
\end{align}
We can observe that:
\begin{align}
    & \left|\sin(i\pi\nu)\right| = \left|\sinh(\pi\nu)\right| = \sqrt{\frac{\cosh\left(2\pi\mathrm{Re}\nu\right)-\cos\left(2\pi\mathrm{Im}\nu\right)}{2}} \ge \\
    & \ge \frac{1}{\sqrt{2}}\sqrt{\frac{1}{2}e^{2\pi\mathrm{Re}\nu}-1} \sim \frac{1}{2}e^{\pi \mathrm{Re}\nu}
\end{align}
Now we use the hypothesis (\ref{eq:hypothesis}) so that:
\begin{align}
    \left|\nu\rho\left(\frac{d}{2}-i\nu\right)g\left(\frac{d}{2}-i\nu\right)\psi_{\frac{d}{2}-i\nu}(\xi)\right| \lesssim \mathcal{A}(\xi) |\nu|^{1/2} [c(\xi)]^{\mathrm{Im}\nu}
\end{align}
where $\mathcal{A}(\xi)$ is another constant.
Therefore:
\begin{align}
    &R\int_0^\pi d\theta \left|\rho\left(\frac{d}{2}-iRe^{i\theta}\right) g\left(\frac{d}{2}-iRe^{i\theta}\right)\psi_{\frac{d}{2}-iRe^{i\theta}}(\xi)\right| \lesssim \tilde{\mathcal{A}}(\xi) \sqrt{R} \int_0^\pi d\theta \left[c(\xi)\right]^{R\sin\theta}=\\
    & =\pi \tilde{\mathcal{A}}(\xi) \sqrt{R} \left[I_0\left(R \log c(\xi)\right)+\mathbf{L}_0\left(R \log c(\xi)\right)\right] \xrightarrow[R \to \infty]{}0
\end{align}
where $I_n$ and $\mathbf{L}_n$ are respectively the modified Bessel function of the first kind and the modified Struve function and $\tilde{\mathcal{A}}(\xi)$ is again a constant.
\section{SK perturbation theory for the 2-point function}\label{app:SK}
In the BEFT one can treat perturbatively both the conservative and non conservative interactions with the SK formalism \cite{Kamenev}. The exception is the $\phi^2$ operator which is relevant and must be treated non-perturbatively. So as the free action we take
\be
S_{0} =\int_{-\infty}^{+\infty} dt \int_0^\infty dx \ \left[\frac{1}{2}\left(\dot\phi^2 -{\phi'}^2\right) -\delta(x) c_0 \phi^2\right].
\ee
The presence of the ``mass'' produces the phase shift:
\be
    e^{2i\alpha(\omega)} = \frac{\omega-i c_0}{\omega+i c_0}
\ee
Differently from the non perturbative approach followed in section (\ref{sec:EFT}), one can treat perturbatively with the SK formalism also the conservative term. Indeed one can define the SK action starting from the action (\ref{EFT}) as:
\be
S_{\rm SK}[\phi_R,\phi_L] = S_{\rm BEFT}[\phi_R]-S_{\rm BEFT}[\phi_L]
\ee
The conservative term in SK language are all of the form $\partial_t^n \phi_+\partial_t^n \phi_-$.

The insertion of a generic derivative operator inside the action of the form:
\begin{align}
\label{eq:genericinteraction}
    S_g[\phi_+,\phi_-]=-g\int_{-\infty}^{+\infty}dt \partial_t^n \phi_+(t,0) \partial_t^m \phi_-(t,0)
\end{align}
gives, at $\mathcal{O}(g)$, the following deformation of the commutator:\footnote{In what follows $\llangle \cdot \rrangle$ means expected value in the SK formalism.} 
\begin{align}
    &\delta_g \left\langle \left[\phi(X),\phi(Y)\right]\right\rangle \equiv \delta_g \left\llangle \phi_+(X)\phi_-(Y)\right\rrangle =\\ &=-ig\int_\mathbb{R} dt \, \partial_t^n \left\llangle \phi_+(t,0)\phi_-(Y) \right\rrangle_0 \partial_t^m \left\llangle \phi_-(t,0)\phi_+(X) \right\rrangle_0,
\end{align}
where the subscript 0 means that those are computed in the free theory. The SK correlator are connected with usual correlators as:
\begin{align}
    & \left\llangle \phi_+(t_X,x)\phi_-(t_Y,y) \right\rrangle_0 = \Theta(t_X-t_Y) \left[\phi(t_X,x),\phi(t_Y,y)\right]_0; \\
    & \left\llangle \phi_-(t_X,x)\phi_-(t_Y,y) \right\rrangle_0 = 0; \\
    & \left\llangle \phi_+(t_X,x)\phi_+(t_Y,y) \right\rrangle_0 = \frac{1}{2}\langle\left\{\phi(t_X,x),\phi(t_Y,y)\right\}\rangle_0; \\
    & \left[\phi(t_X,x),\phi(t_Y,y)\right]_0 = -\frac{i}{\pi}\int_{-\infty}^{+\infty}\frac{d\omega}{\omega}\cos\left(\omega x+\alpha(\omega)\right) \cos\left(\omega y+\alpha(\omega)\right) \sin\left[\omega(t_X-t_Y)\right];\\
    & \left\langle\left\{\phi(t_X,x),\phi(t_Y,y)\right\}\right\rangle_0 = \frac{2}{\pi}\int_{-\infty}^{+\infty}\frac{d\omega}{\omega\left(e^{2\pi\omega}-1\right)}\cos\left(\omega x+\alpha(\omega)\right) \cos\left(\omega y+\alpha(\omega)\right) \cos\left[\omega(t_X-t_Y)\right].
\end{align}
The non-conservative terms inside the BEFT action appear as in eq. (\ref{eq:dissipative}). With these tools we can compute the effect at linear level in the coupling of both the conservative and non-conservative terms:
\begin{align}
    & \delta_{\rm c}\left\langle\left[\phi_{\rm out}(u),\phi_{\rm in}(0)\right]\right\rangle =-i\sum_{n=1}^\infty c_n \int_{-\infty}^{+\infty}\frac{d\omega}{2\pi} \, \omega^{2(n-1)}\cos^2\alpha(\omega) e^{2i\alpha(\omega)-i\omega u} \\
    & \delta_{\rm nc}\left\langle\left[\phi_{\rm out}(u),\phi_{\rm in}(0)\right]\right\rangle =-\sum_{n=0}^\infty b_n \int_{-\infty}^{+\infty}\frac{d\omega}{2\pi} \, \omega^{2n-1}\cos^2\alpha(\omega) e^{2i\alpha(\omega)-i\omega u}
\end{align}
This means, if we expand in powers of $\omega$ the EFT version of the function that appears in (\ref{Comint}):
\begin{align}
    \mathcal{S}^{\rm EFT}(\omega) = -1-\frac{2i}{c_0}\omega+\frac{2(b_0+1)}{c_0^2}\omega^2+\mathcal{O}\left(\omega^3\right).
\end{align}
To achieve the $\mathcal{O}(\omega^3)$ term in the above expansion we need to include not only the conservative interaction at linear order $-c_1 \dot{\phi}^2(t,0)$, but also the dissipative interaction $-b_0 \left(\phi_- \dot{\phi}_+\right) (t,0)$ at quadratic level $\mathcal{O}\left(b_0^2\right)$.

From the path integral formulation one can see that:
\begin{align}
    &\langle\left[\phi(X),\phi(Y)\right]\rangle=\left\llangle \phi_+(X)\phi_-(Y) \right\rrangle = \\
    &=\left\llangle \phi_+(X)\phi_-(Y) \right\rrangle_0+\underbrace{i\left(\left\llangle \phi_+(X)\phi_-(Y) S_{\rm int} \right\rrangle_0-\left\llangle \phi_+(X)\phi_-(Y)\rrangle_0\llangle S_{\rm int} \right\rrangle_0\right)}_{\mathcal{O}\left(b_0\right)}+\\
    \label{eq:secondordercontribution}
    & \left.\begin{array}{lr}-\frac{1}{2}\left\llangle \phi_+(X)\phi_-(Y) S^2_{\rm int} \right\rrangle_0+\left\llangle \phi_+(X)\phi_-(Y) S_{\rm int} \right\rrangle_0 \left\llangle S_{\rm int} \right\rrangle_0+\\
    +\left\llangle \phi_+(X)\phi_-(Y)\right\rrangle_0\left(\frac{1}{2}\left\llangle S^2_{\rm int} \right\rrangle_0-\left\llangle S_{\rm int} \right\rrangle_0^2\right)+\end{array}\right\}\mathcal{O}(b_0^2)\\
    &+\mathcal{O}\left(b_0^3\right).
\end{align}
For the specific case of $S_{\rm int}=-b_0\int_\mathbb{R}dt\dot{\phi}_+(t,0) \phi_-(t,0)$, after the Wick's contraction, we have that the $\mathcal{O}(b_0^2)$ term is:
\begin{align}
    \delta_{b_0^2}\langle\left[\phi(X),\phi(Y)\right]\rangle=-b_0^2\int_{\mathbb{R}^2}dt dt' \left\llangle \phi_+(X) \phi_-(t,0) \right\rrangle_0 \left\llangle \dot{\phi}_+(t,0) \phi_-(t',0) \right\rrangle_0 \left\llangle \dot{\phi}_+(t',0) \phi_-(Y) \right\rrangle_0.
\end{align}
So eventually we have that:
\begin{align}
    \delta_{b_0^2}\langle\left[\phi_{\rm out}(u),\phi_{\rm in}(0)\right]\rangle=\int_{-\infty}^{+\infty}\frac{d\omega}{2\pi} e^{-i\omega u}\left[\frac{ib_0^2}{c_0^3}\omega^2+\mathcal{O}\left(\omega^3\right)\right],
\end{align}
from which:
\begin{align}
\label{SEFT}
    \mathcal{S}^{\rm EFT}(\omega) = -1-\frac{2i}{c_0}\omega+\frac{2(b_0+1)}{c_0^2}\omega^2+\frac{2i(1+2b_0+b_0^2+c_0c_1)}{c_0^3}\omega^3+\mathcal{O}\left(\omega^4\right).
\end{align}
As a consistency check, setting $b_0 =0$ gives the expansion of the exact phase shift \eqref{tan} up to order $\omega^3$. 

One might go beyond the quadratic action including a $\phi^3$ interaction. The leading contribution in (\ref{SEFT}) is at $\mathcal{O}(\omega)$, that is the same order at which dependence on $c_0$ shows up.
\bibliography{bibstat}

\end{document}